\documentclass[a4paper,11pt]{article}
\usepackage{jinstpub} 
\usepackage{lineno}

\title{\boldmath Characterisation of a low-momentum high-rate muon beam monitor for the FAMU experiment at the CNAO-XPR beam facility}

\author[a,b,c,*]{R. Rossini,\note[*]{Corresponding author.}}
\author[d,l]{R. Benocci,}
\author[d]{R. Bertoni,}
\author[d,e]{M. Bonesini,}
\author[d,f]{S. Carsi,}
\author[d,e]{M. Clemenza,}
\author[b]{A. de Bari,}
\author[g,h]{M. Donetti,}
\author[b]{C. de Vecchi,}
\author[a,b]{A. Menegolli,}
\author[g,i]{A. Mereghetti,}
\author[j]{E. Mocchiutti,}
\author[d,f]{C. Petroselli,}
\author[b]{M.C. Prata,}
\author[b,g]{M. Pullia,}
\author[b]{G.L. Raselli,}
\author[b]{M. Rossella,}
\author[g,i]{S. Savazzi,}
\author[k]{L. Tortora,}
\author[d]{E.S. Vallazza,}

\affiliation[a]{Dept. of Physics, University of Pavia, Pavia, Italy}
\affiliation[b]{Sezione di Pavia, Istituto Nazionale di Fisica Nucleare (INFN), Pavia, Italy}
\affiliation[c]{ISIS Neutron and Muon Source, Science and Technology Facilities Council (STFC), Didcot, UK}
\affiliation[d]{Sezione di Milano-Bicocca, Istituto Nazionale di Fisica Nucleare (INFN), Milan, Italy}
\affiliation[e]{Dept. of Physics "G. Occhialini", University of Milano-Bicocca, Milan, Italy}
\affiliation[f]{Dept. of Science and High Technology, University of Insubria, Como, Italy}
\affiliation[g]{Centro Nazionale di Adroterapia Oncologica (CNAO), Pavia, Italy}
\affiliation[h]{Sezione di Torino, Istituto Nazionale di Fisica Nucleare (INFN), Turin, Italy}
\affiliation[i]{Sezione di Milano, Istituto Nazionale di Fisica Nucleare (INFN), Milan, Italy}
\affiliation[j]{Sezione di Trieste, Istituto Nazionale di Fisica Nucleare (INFN), Trieste, Italy}
\affiliation[k]{Sezione di Roma Tre, Istituto Nazionale di Fisica Nucleare (INFN), Rome, Italy}
\affiliation[l]{Dept. of Earth and Environmental Sciences DISAT, University of Milano-Bicocca, Milan, Italy}

\emailAdd{riccardo.rossini@infn.it}

\abstract{The FAMU experiment aims at an indirect measurement of the Zemach radius of the proton. The measurement is carried out on muonic hydrogen atoms ($\upmu$H) produced through the low-momentum (50-60 MeV/c) muon beam a the RIKEN-RAL $\upmu^-$ facility. The particle flux plays an important role in this measurement, as it is proportional to the number of $\upmu$H atoms produced, which is the target of the FAMU experimental method. The beam monitor calibration technique and results, presented here, are meant to extract a reliable estimation of the muon flux during the FAMU data taking. These measurements were carried out at the CNAO synchrotron in Pavia, Italy, using proton beams and supported by Monte Carlo simulation of the detector in Geant4.}


\begin{document}
\maketitle
\flushbottom

\section{Introduction}\label{sec:intro}
    The FAMU experiment aims at an indirect estimation of the proton Zemach radius by measuring the hyperfine splitting in the ground state of muonic hydrogen\cite{pizzolotto2020}. The experiment is carried out at the RIKEN-RAL Port1 negative muon beamline at the Rutherford Appleton Laboratory (RAL), in Didcot, UK. The observable of the experiment, i.e. the amount of muonic oxygen characteristic X-rays, depends on the number of muonic hydrogen atoms in the target, which is a function of the muonic flux. It is therefore crucial to have an on-line beam monitor to normalise data with the muon flux.

    The negative muon beam used for the FAMU experiment at RIKEN-RAL Port 1 consists in two muon spills delivered with an average rate of 40 Hz. The muon flux in the 50-60 MeV/c momentum range, which contains the FAMU momentum setting, is $ \sim 7 \cdot 10^4$ muons per second\cite{matsuzaki2001, hillier2019}. 

    A set of beam monitors has been developed in the recent years to match the requirements of the FAMU experiment\cite{bonesini2017, bonesini2018}. Such detectors, generally called \emph{hodoscopes}, are made of $32 \times 32$ plastic scintillating fibres read out by SiPMs. To enable the estimation of the muon flux, preliminary calibrations with cosmic muons have been made. However, cosmic muons deposit a different energy with respect to FAMU muons. A new calibration method based on single particle measurement has been proposed on the FAMU hodoscopes\cite{rossini2023_1, rossini2023_2} and the results are discussed here. The calibration, carried out with a low-rate proton beam, took place at the CNAO synchrotron in Pavia, Italy, described in Section \ref{fig:cnao}\cite{rossi2015}. 

\section{Beam hodoscope}\label{sec:hodo}
    The detector tested in the data taking is an upgraded version of the 1 mm pitch beam hodoscope initially designed for the R582 FAMU run\cite{bonesini2017, bonesini2018}. The core of the detector consists of two planes of 1 mm pitch square scintillating fibres (Bicron BCF12). Fibres are coated with a layer of Extra-Mural Absorber (EMA), to reduce optical cross-talk, and cut at a length of 32 mm. Each plane has 32 parallel fibres and the two planes are juxtaposed with crossing fibres in order to enable X-Y identification for a particle entering the detector, forming a fiducial area of $32 \times 32$ mm$^2$. Each fibre is read-out by one end by an Advansid RGB SiPM (active area $1 \times 1$ mm$^2$, cell size 40 $\upmu$m). Signals are fanned out through MCX connectors and acquired using two CAEN V1742 digitisers (32 channels, 5 GS/s, 12 bit).

\section{Calibration technique}\label{sec:calib}
    The idea behind the calibration technique has already been presented in previous works\cite{rossini2023_1, rossini2023_2}. In general, the problem can be summarised as follows:
    \begin{itemize}
        \item cosmic muons have a low rate, but they are mainly MIPs (Minimum Ionising Particles). The energy loss $dE/dx$ of MIPs in the hodoscope is much smaller than the one of FAMU muons. Thus, a calibration performed with cosmic rays introduces a systematic uncertainty resulting from the different values of $dE/dx$;
        \item available muon sources at the desired momentum are unable to reach rates low enough to have single particle signals.
    \end{itemize}
    
    \begin{figure}[htbp]
        \centering
        \includegraphics[width=\textwidth]{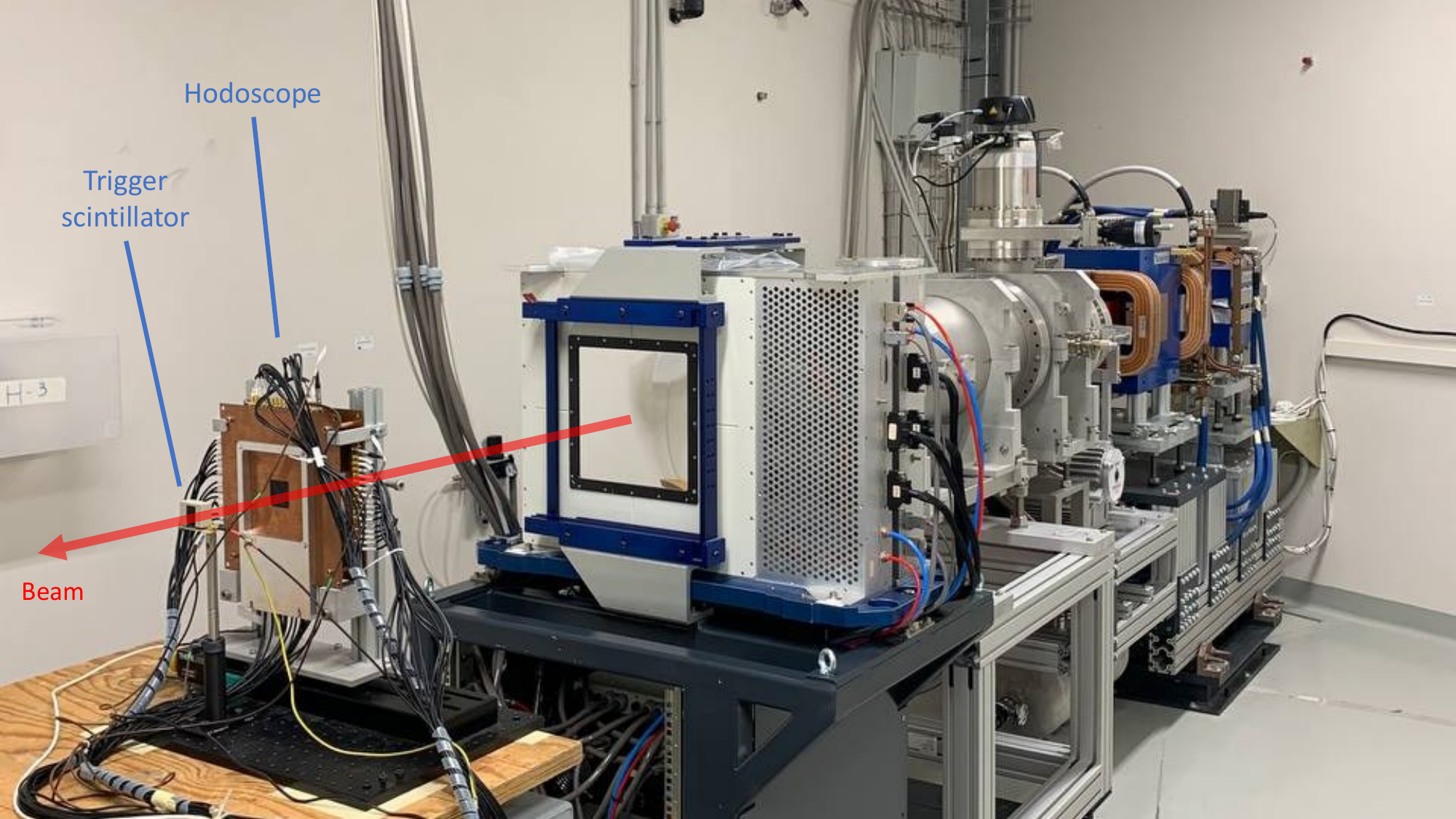}
        \caption{The beam hodoscope characterised in this work mounted at the CNAO-XPR beamline.\label{fig:setup}}
    \end{figure}
    
    The solution proposed is to calibrate the detector with proton beams, much more available even at low rate), selecting the energy in order to match the $dE/dx$ of FAMU muons. Thorough Monte Carlo simulation of the detector in Geant4 show that such protons have total energy of 150 MeV (momentum 550 MeV/c). A low-rate 150 MeV proton beam is a reachable condition at CNAO in treatment and experimental rooms, where the measurements reported here were carried out. Figure \ref{fig:setup} shows the hodoscope mounted at the experimental beamline of the CNAO synchrotron, better described in Section \ref{sec:cnao}. 

    A cubed plastic scintillator (1 cm$^3$), positioned downstream from the hodoscope, was used to trigger the acquisition system. The beam was tuned to have a rate low enough to acquire single particles on each trigger.

\section{The CNAO-XPR facility}\label{sec:cnao}
\begin{figure}[htbp]
        \centering
        \includegraphics[width=\textwidth]{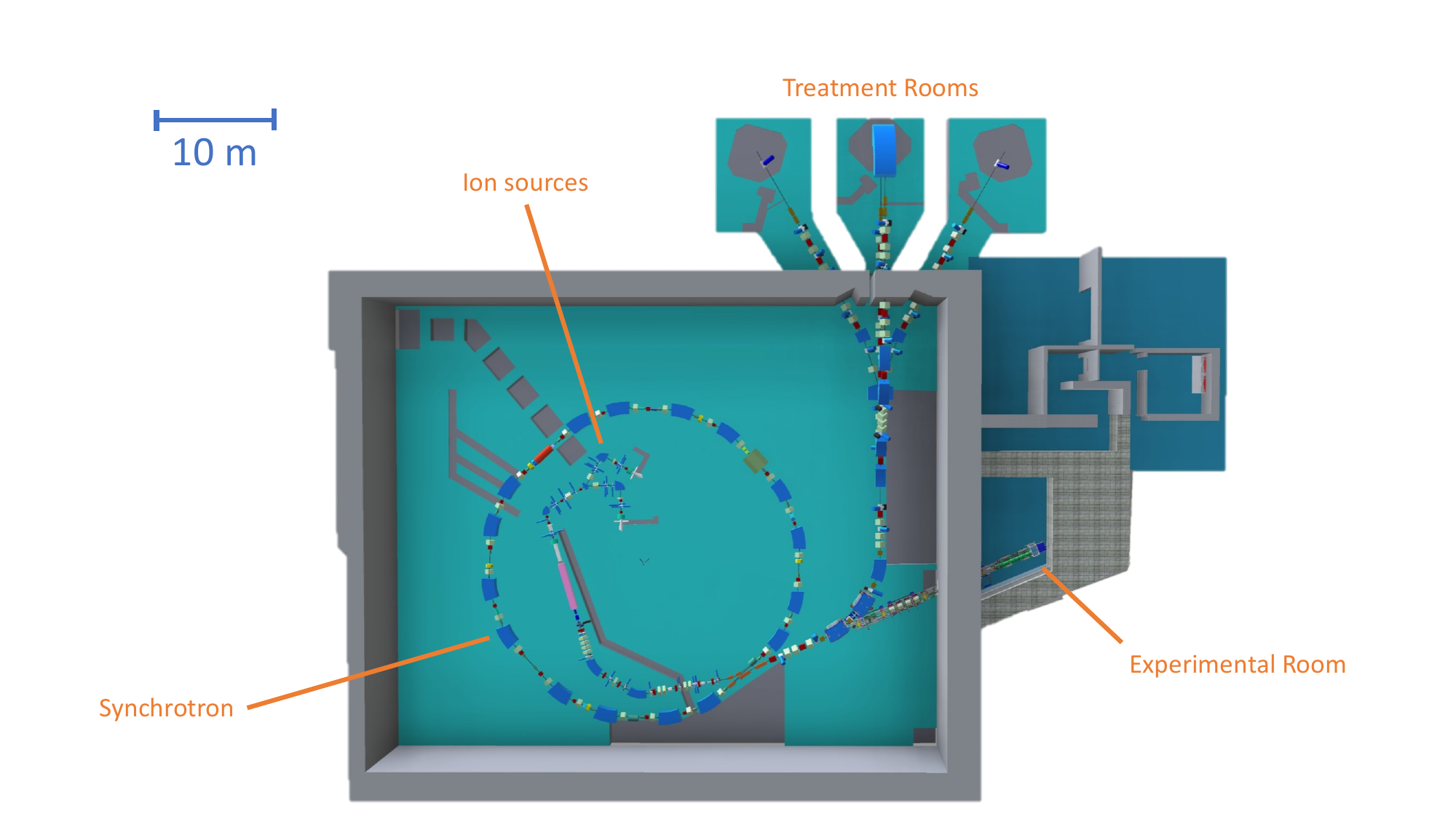}
        \caption{The CNAO synchrotron: ion sources and position of experimental and treatment halls.\label{fig:cnao}}
    \end{figure}

    Centro Nazionale di Adroterapia Oncologica (CNAO, \emph{Italian National Centre for Oncologic Hadron Therapy}), in Pavia (Italy), is one of the four centres in Europe, and six worldwide, offering treatment of tumours with both protons and carbon ions. Besides clinical activity, CNAO has also \emph{research} as institutional purpose. For this reason, in addition to the three treatment rooms, the CNAO center is equipped with an \emph{experimental room} (CNAO-XPR) dedicated to experimental activities and available also to external researchers. Typical research activities include radiation biophysics, radiobiology, space research, materials research nuclear physics and detector development. A schematic illustration of the CNAO accelerator and beamline layout is shown in Figure \ref{fig:cnao}.
    
    The CNAO synchrotron provides energies up to 400 MeV/u for carbon ions (corresponding to a Bragg peak depth of up to 27 cm in water) and up to 227 MeV for protons (corresponding to a Bragg peak depth of up to 32 cm in water). The minimum extraction energies are about 115 MeV/u and 63 MeV, for carbon and protons respectively, corresponding to particle range of 30 mm. An additional ion source was recently installed that will provide additional ion species (He, Li, O and Fe) for research and possibly for clinics in a second stage. Its commissioning will start at the beginning of 2024. Furthermore, access to a biological laboratory with all the necessary equipment including laminar flows, incubators, centrifuges, cell counters is available on site.

    The CNAO beam is extracted from the synchrotron by a “resonant slow extraction”, to distribute the extracted particles over a time of about 1 s (called a spill). Between the end of a spill and the beginning of the following one, the accelerator magnets are ramped up to the maximum field, then to the minimum field and finally brought back to the injection value from which a new beam can be accelerated. This leads to a discontinuous time structure of the beam with a pause of a couple of seconds between one spill and the following one. The beam energy can be changed at each spill. The intensity of the beams available in the experimental room are the same as those used for the medical treatments and therefore nominal intensities of the order of $10^{10}$ protons/s and $4 \cdot 10^8$ carbon ions/s. The minimum intensities are one or two orders of magnitude lower. For many applications, as it was the case for the characterization of the FAMU hodoscope, much lower intensities are required; they can be achieved with appropriate accelerator settings, but in this case the intensity is too low for the beam monitoring system which is practically blind. In these cases, it is up to the experimenter to give feedback to the control room for attaining the necessary beam intensity for his experiment. However, a suitable beam monitor for low intensities is under development; once operational, it will be installed in the experimental room at needs. The minimum intensities reachable are in the order of hundreds of particles per second. The beam is about one centimetre in diameter, as an order of magnitude. It is possible to scan a significantly larger area thanks to a system of deflector magnets that move the beam transversely both horizontally and vertically, as it happens for treatments. According to the needs of the experiment to carry out, the experimental beamline can be arranged in four different configurations depending on the space required downstream the target or the dimensions of the scanning field. The maximum area that can be covered is 20 cm x 20 cm. The positions of the 4 irradiation positions (called in jargon \emph{isocentres}) can be visualized during experiment installation thanks to a system of lasers lines, similar to those projected by a laser level. The projected planes intersect at the four irradiation points and are directed in the horizontal plane, in the sagittal plane (vertical along the beam) and in the transverse plane (perpendicular to the beam).

\section{Data analysis and results}\label{sec:results}
    The first step of data processing consisted in extracting, for every \emph{event} (i.e. for every trigger), the pulse integral over time of each fibre. As a SiPM signals consists in a current pulse $I_{fib}(t)$, its integral is a total deposited charge $Q_{fib}$, which is proportional to the total energy deposited by the particle in the fibre $\Delta E_{fib}$. These pulse integrals were then processed to select single-particle events only. 
    
\begin{figure}[htbp]
        \centering
        \includegraphics[width=\textwidth]{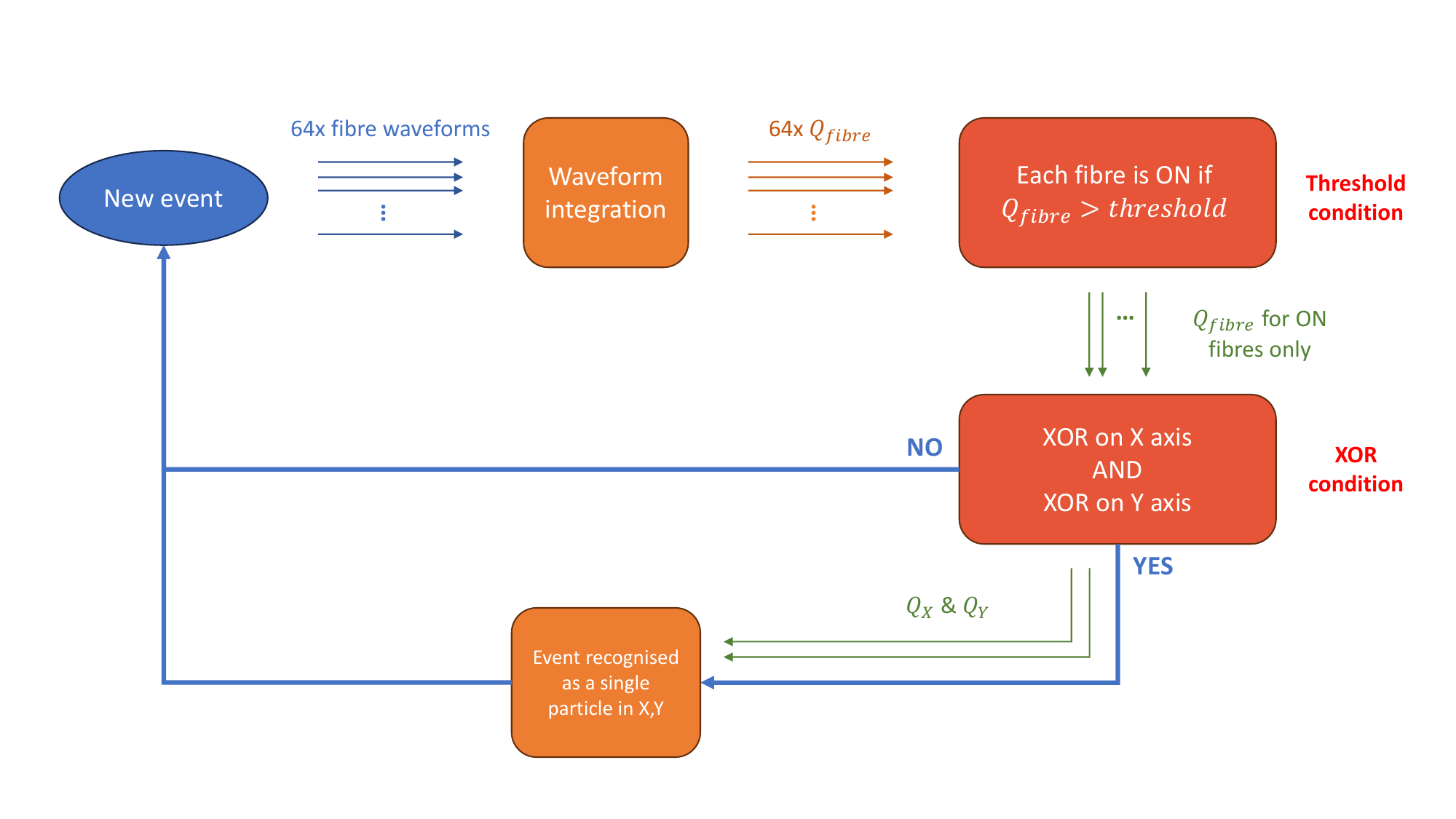}
        \caption{Schematic view of the waveform processing for data analysis.\label{fig:processing}}
    \end{figure}

\begin{figure}[htbp]
        \centering
        \includegraphics[width=.8\textwidth]{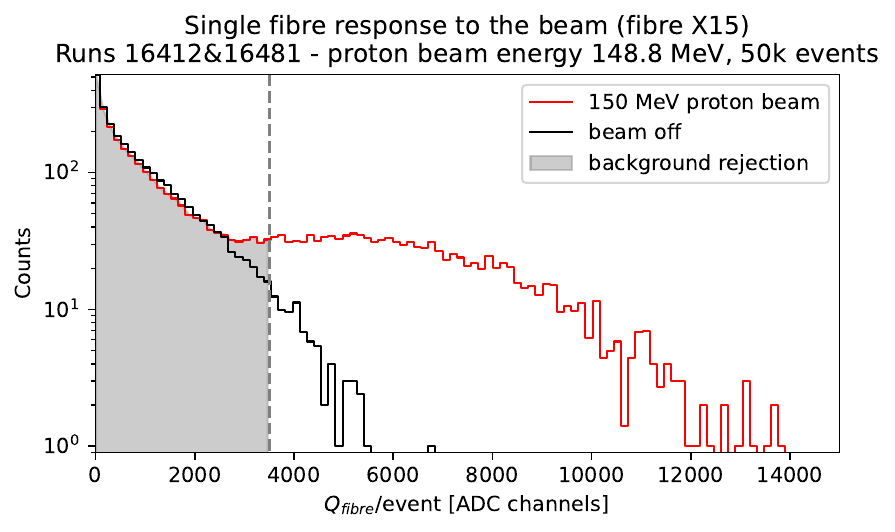}
        \caption{Response of a single fibre (X15 in this case) immersed into the beam compared to a background acquisition with the same statistics (50k events). As shown, background rejection for a single fibre consists in setting a threshold in $Q_{fibre}$ for each event and each fibre. The threshold shown in this plot (3500 ADC channels) is the value used for this analysis, coming from threshold optimisation (see Figure \ref{fig:threshold}.\label{fig:Fib}}
    \end{figure}

    In particular, for each event, two conditions are applied (see Figure \ref{fig:processing}) as follows. The first step is to decide whether a fibre is to be considered ON or OFF on that event. This is carried out by imposing a threshold \texttt{thrs} on $Q_{fib}$. As an example, Figure \ref{fig:Fib} is a histogram of values of $Q_{fib}^{X15}$ (for fibre X15) with and without beam. This condition considers fibre X15 ON when $Q_{fib}^{X15} > \mathtt{thrs}$, as such signals are considered caused by the passage of a particle in fibre X15. This first step is referred to as the \emph{threshold condition}.
    
    After the first step, which basically defines whether a fibre is ON or OFF, we want to select single-particle events only. In the easiest approximation, particles are expected to cross one fibre for each plane and exit the detector on the other side. Thus, single-particle discrimination is obtained by checking for each event if exactly two fibres are ON, and in particular they must be one on the $X$-oriented plane and one on the $Y$-oriented plane. As the strongest condition here is the exclusive \texttt{OR} (\texttt{XOR}) on each fibre plane, this condition is referred to as the \emph{XOR condition}.

\begin{figure}[htbp]
        \centering
        \includegraphics[width=.8\textwidth]{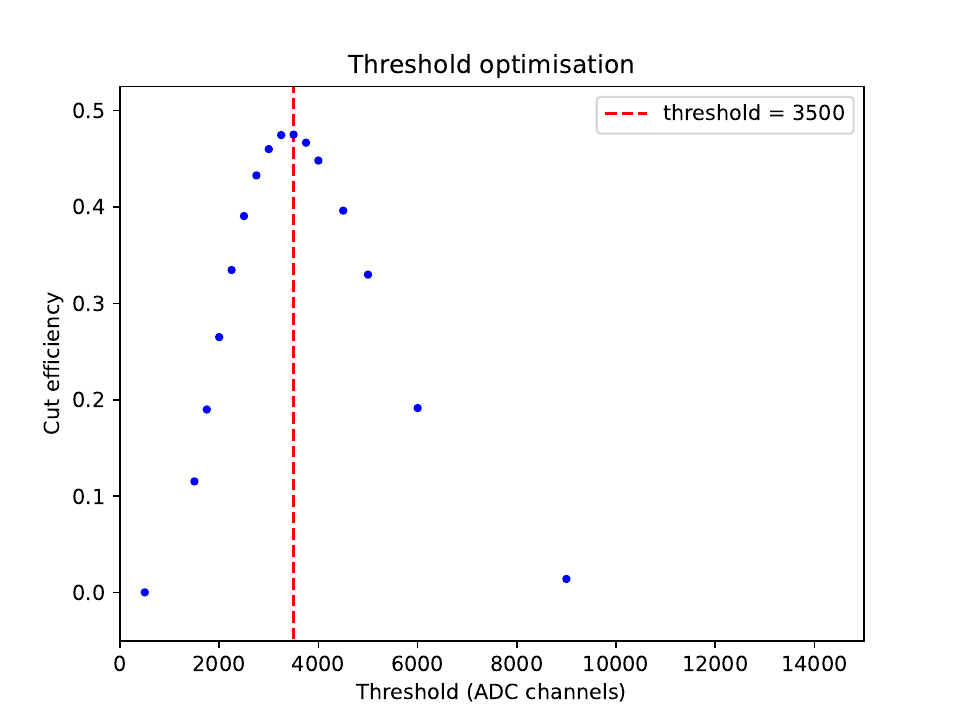}
        \caption{Optimisation plot of the threshold for 150 MeV proton beam. For low threshold, the number of fibres considered ON increases and the XOR condition fails consequently. For high threshold, fibres are considered ON only a few times due to common failing of the threshold condition.  \label{fig:threshold}}
    \end{figure}

    These two conditions can be summed up with the following logic equation ($\oplus = \mathtt{XOR}$):
    \begin{equation}
        \forall \ \text{event, } \ \text{if }\left( \bigoplus_{j=1}^{32} Q_{fib}^{Xj} > \texttt{thrs} \right) \mathtt{AND} \left( \bigoplus_{k=1}^{32} Q_{fib}^{Yk}  > \texttt{thrs} \right) \ \Rightarrow \ \text{single-particle event}
    \end{equation}
    where it is clear that the number of selected events strongly depends on the value chosen for the threshold. In fact (see Figure \ref{fig:Fib}): 
    \begin{itemize}
        \item if $\mathtt{thrs} \to 0$, the threshold condition becomes too lose. In this case, fibres are considered ON even with background signals and therefore the XOR condition rejects almost all events; 
        \item if $\mathtt{thrs} \to \max(Q_{fib})$, the threshold condition becomes too strict and fibres with signals are considered OFF. In this case, it is the threshold condition rejecting all events.
    \end{itemize}
    As a consequence, there exists an optimum value of threshold which rejects most of the background without impacting too much the true signal. For the purpose of this study, it has been decided to set a threshold which maximises the efficiency of the selection carried out by the above conditions. The optimisation plot, from which a threshold of 3500 ADC channels was chosen, is shown in Figure \ref{fig:threshold}. The plot in Figure \ref{fig:Fib}, already mentioned above, uses the optimised threshold (grey dashed line) to discriminate the signal from the background.

\begin{figure}[htbp]
        \centering
        \includegraphics[width=\textwidth]{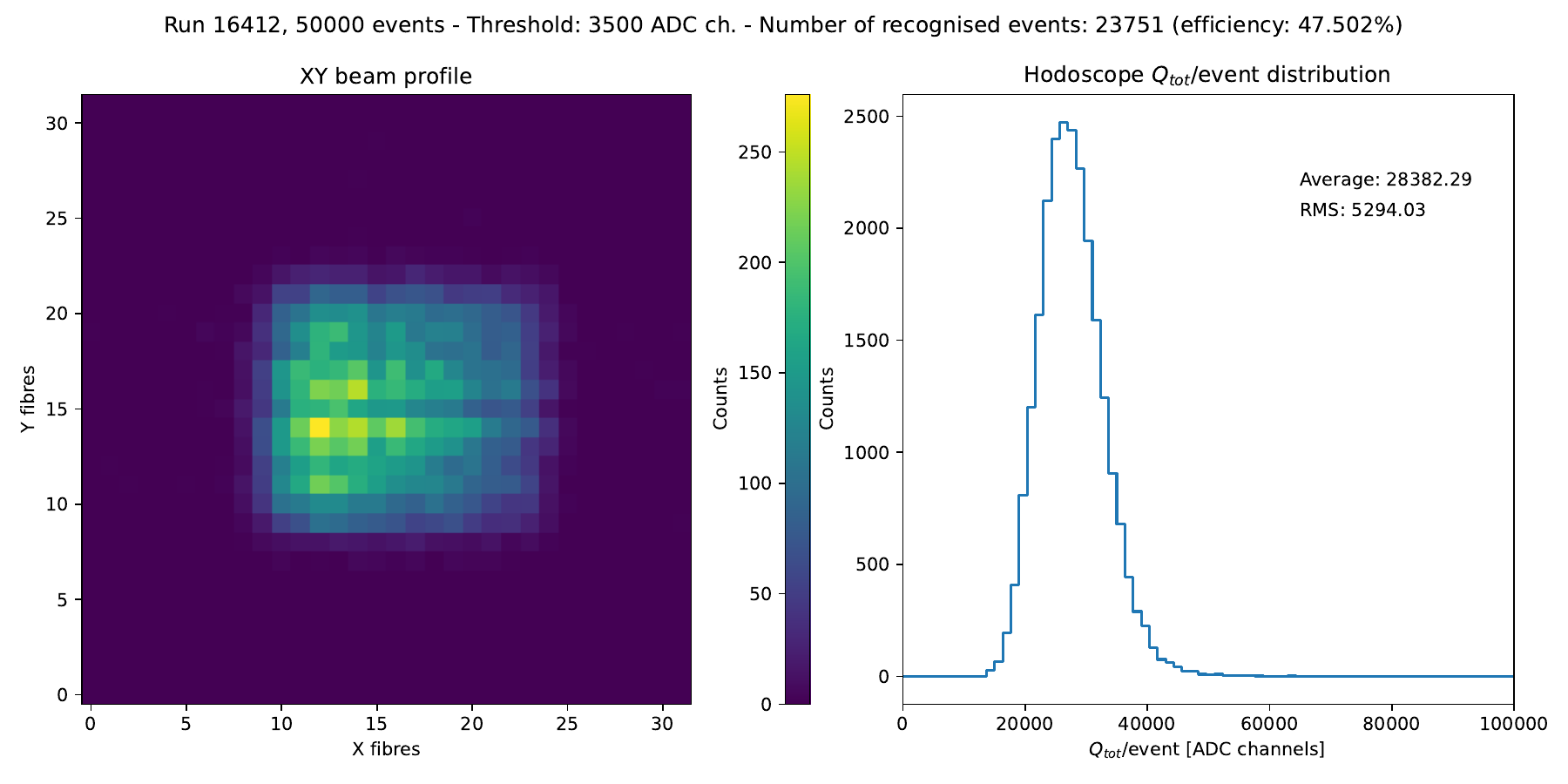}
        \caption{Threshold at 3500 ADC channels, 150 MeV reconstructed proton beam shape (left) and histogram of deposited charge $Q_{tot}$ per event in the detector (right).\label{fig:spectra}}
    \end{figure}

    Figure \ref{fig:spectra} shows the $XY$ beam map and the histogram of total deposited charge $Q_{tot}$ with the optimised threshold. The $XY$ beam map is obtained by considering the crossing of the two fibres considered ON in the events selected by the conditions. The beam profile appears asymmetric (on the $X$ axis only) and squared because of the shape of the trigger detector and the fact it had probably been misaligned on the $X$ axis by a few mm with respect to the beam isocentre. The $Q_{tot}$ histogram is obtained summing all deposited charges for the events selected ($Q_{tot} = \sum_{j=1}^{32} (Q_{fib}^{Xj}+Q_{fib}^{Yj})$). The value of equivalent deposited charge for these protons ($Q_p$), which is equal to the one of FAMU muons ($Q_\mu$) for the choice of proton energy, is obtained as the mean of a gaussian fit around the peak of the $Q_{tot}$ distribution: $Q_\mu = (27.95 \pm 0.05) \cdot 10^3$ ADC channels. At the RIKEN-RAL facility, when detecting the stream of muons in a beam spill, the total integrated charge $Q$ can be converted into the number of muons with the equation $N = Q/Q_\mu$, where $Q_\mu$ comes from the calibration procedure described above.

\section{Summary}\label{sec:conclusion}
In conclusion, the calibration procedure described in \cite{rossini2023_1, rossini2023_2} has been carried out on a detector made of two planes of 32 adjacent scintillating fibres having 1 mm pitch. The measurements have been carried out at the CNAO synchrotron and the conditions applied during the analysis allowed to select the single-particle events. The optimisation of the threshold on which conditions are based on has been discussed, as it is a crucial point for trusting the analysis protocol. Finally, the value of $Q_\mu = (27.95 \pm 0.05) \cdot 10^3$ ADC channels has been obtained and will be used for the calculation of the incoming flux at the RIKEN-RAL muon facility.

\acknowledgments
These results were obtained also thanks to the use of the CNAO experimental facility built in collaboration with INFN.




\bibliographystyle{JHEP}
\bibliography{biblio.bib}






\end{document}